\documentstyle[prb,aps,graphicx]{revtex}

\begin{document}
\draft
\preprint{Phys.Rev. B 63, 205118 (2001) }
\title{Symmetry crossover and excitation thresholds at the 
neutral-ionic transition of the modified Hubbard model}

\author{Y. Anusooya-Pati and Z.G. Soos}
\address{Department of Chemistry, Princeton University, NJ-08544 Princeton US}

\author{ A. Painelli}
\address{Dipartimento Chimica G.I.A.F., Universita' di Parma, I-43100 Parma, 
Italy}

\date{\today}
\maketitle

\begin{abstract}

Exact ground states, charge densities and excitation energies are 
found using valence bond methods for $N$-site modified Hubbard models 
with uniform spacing. At the neutral-ionic transition (NIT), the ground 
state has a symmetry crossover in 4n, 4n+2 rings with periodic and
 antiperiodic boundary conditions, respectively. Large site 
energies $\Delta$ stabilize a paired state of the half-filled chain,
 while large $U$ stabilizes a covalent state. Finite transfer 
integrals $t$ shift the NIT to the covalent side of $U - 2\Delta$. Exact
 results to $N$ = 16 in the full basis and to $N$ = 22 in a restricted 
basis for large $U$, $\Delta$ are extrapolated to obtain the crossover and 
charge density of extended chains. The modified Hubbard model has 
a continuous NIT between a diamagnetic band insulator on the paired
 side and a paramagnetic Mott insulator on the covalent side. 
The singlet-triplet (ST), singlet-singlet (SS) and charge gaps for finite 
$N$ indicate that the ST and SS gaps close at the NIT with increasing $U$ 
and that the charge gap vanishes only there. Finite-$N$ excitations constrain 
all singularities to $\pm$  0.1$t$ of the symmetry crossover. The NIT is 
interpreted as a localized ground state (GS) with finite gaps on 
the paired side and an extended GS with vanishing ST and SS gaps 
on the covalent side. The charge gap and charge stiffness indicate
 a metallic GS at the transition  that, however, is unconditionally 
unstable to dimerization. Finite $\Delta$ breaks electron-hole (e-h) symmetry, 
but the modified Hubbard model has an extended e-h symmetry and strong
 mixing of spin and charge excitations is limited to a few $t$'s about the NIT.
 Exact finite-size results complement other approaches to 
 valence or ferroelectric transitions 
in organic charge-transfer salts or in inorganic oxides, and to  
electron-vibration coupling and structural instabilities in 
one-dimensional systems.

\end{abstract}
\narrowtext
\twocolumn

\section{Introduction}

McConnell and coworkers\cite{mcconnell} explained the sharp separation
 of organic charge transfer (CT) complexes into diamagnetic and paramagnetic
 by proposing that weak $\pi$-donors (D) and acceptors (A) 
form neutral complexes of molecules, while strong donors and 
acceptors crystallize as ion radicals D$^+$ and A$^-$. 
These planar conjugated systems form one-dimensional structures, 
either as mixed ...DADA... stacks in CT complexes or as segregated stacks 
in ion-radical salts \cite{soosklein}.  
The simplest approximation for a crossover between DA and D$^+$A$^-$ 
ground states is $M = E_I - E_A$, where $M$ is the Madelung energy, 
$E_I$ is the ionization potential of the donor and $E_A$ is the electron 
affinity of the acceptor. Although $M$ is inherently long ranged, 
the systems are quasi one-dimensional by virtue of $\pi$-overlap 
restricted to stacks. Strebel and Soos \cite{strebelsoos} introduced 
the modified Hubbard model with transfer integral 
$t = -\langle DA|H|D^+A^-\rangle $ for CT complexes and studied the 
crossover in the random phase approximation. 
Finite $t$ leads to mixing and to partial ionicity D$^{\rho +}$A$^{\rho -}$ 
in the ground state (GS), with $\Delta \rho < 1$ at the crossover. 
The modified Hubbard model, Eq.~(\ref{uno}) 
below, has proved to be extremely rich and widely applicable. 
It describes any valence transition, is a special case of 
important solid-state models, and provides the starting point 
for electron-phonon (e-ph) coupling. Its scope is still growing and 
attracting new theoretical and computational approaches to the 
crossover region, $M \sim E_I - E_A$. We present in this paper exact 
solutions of finite-size systems, including low-lying excitations.

The neutral-ionic transition (NIT) originates with the TTF-CA 
complex studied by Torrance and coworkers \cite{torrance},
with D = tetrathiafulvalene and A = chloranil. 
TTF-CA is neutral at room temperature, with $\rho \sim$ 0.3, 
and has a transition at T $\sim$ 81K to an ionic state with 
$\rho \sim$ 0.7 that, moreover, is dimerized. 
The uniform TTF-CA spacing above 81K becomes alternating  ($...t_1,t_2...$)
in the ionic phase, and partial ionicity is determined 
spectroscopically \cite{girlando}.
 The structural change shows the fundamental role of lattice 
phonons and the Peierls instability of the paramagnetic phase. 
The alternating phase is potentially ferroelectric and the system 
may be metallic at the NIT. Such features are common to
 ferromagnetic oxides, and in this context they
have recently been discussed \cite{egami} in terms of the modified 
Hubbard model. The interplay of electron-electron (e-e) and e-ph 
interactions can generate either continuous or discontinuous 
ionicity changes. Long-range Coulomb interactions can 
generate \cite{soosmazu,gp86,pg88} discontinuous $\rho$ 
variations at the NIT 
as well as strongly affect \cite{pg88} the dimerization instability.

Rice \cite{rice} pointed out the strong infrared activity of 
totally symmetric  molecular vibrations through coupling 
to charge fluctuations. These on-site (Holstein) phonons also 
participate in the NIT. They condense at the transition 
and produce discontinuous  $\rho $ variations  above a critical coupling 
strength \cite{pg88}. 
Electron-molecular-vibration coupling provides the basis for
the spectroscopic  determination of the ionicity, $\rho$, 
of the  D$^{\rho+}$A$^{\rho -}$  GS as well as of the local symmetry,
making vibrational spectroscopy an useful tool to follow
charge and structural phase transitions \cite{bibbie}. 
Joint theoretical and experimental analysis of electronic and 
vibrational spectra \cite{pg87} allowed for a systematic 
characterization of several salts \cite{sistematica}.

The modified Hubbard model adds site energies $\pm \Delta$
to a Hubbard chain with uniform spacing,

\begin{equation}
  H_{0}(t,\Delta ,U) =  
-\sum_{i,\sigma} \left\{ t \left( a^+_{i,\sigma}a_{i+1,\sigma}+
 a^+_{i+1,\sigma}a_{i,\sigma} \right)
-\Delta (-1)^i  a^+_{i,\sigma}a_{i,\sigma} \right\}
+ \sum_i U a^+_{i,\alpha} a^+_{i,\beta} a_{i,\beta}a_{i,\alpha}
\label{uno}
\end{equation}
D and A are at odd and even $i$, respectively,  in the context of 
CT complexes, and $t >0$, $\Delta \geq $ 0 can be 
taken without loss of generality. 
The half-filled case, with one electron per site, is by far the 
most important. We consider $H_0$ at this filling for uniform $t$ 
and equal $U \geq $ 0 for donors and acceptors. 
The electron density on D sites is related to the GS energy,

\begin{equation}
n_D-1 = -\frac{1}{N} \frac{\partial E_0(t,\Delta , U)}{\partial \Delta}
\label{due}
\end{equation}
The ground states of extended systems are not known exactly 
for $\Delta \neq 0$. Approximate solutions beyond mean field have 
been proposed along several lines:
 exact diagonalization \cite{soosmazu,gp86,pg88,pg92},
quantum Monte Carlo \cite{nagaosa}, renormalization group 
methods\cite{avignon}, and continuum models \cite{horovitz}.
There is broad agreement as well as open or disputed points mentioned below.

Hubbard models are readily generalized and provide a unified 
approach to quantum cell models that need not be low-dimensional. 
In the context of Eq.~(\ref{uno}), we note that $\Delta$
 can incorporate the Madelung energy in a mean-field 
approximation \cite{strebelsoos} or coupling to Holstein phonons 
in the adiabatic approximation \cite{pg88,pg92}.
 In either case, the effective $\Delta$ depends on the GS ionicity 
and the NIT becomes discontinuous above a critical coupling. 
The model is then nonlinear and has wider applications to 
susceptibilities \cite{nlo}. At $\Delta = 0$, if $t$ is linearly 
expanded around the equilibrium bond-length, we get a  
Peierls-Hubbard model \cite{torino}, 
and a direct connection to models for ion-radical salts such as TTF-TCNQ 
with segregated stacks.
Alternating transfer integrals $t(1\pm \delta)$ are found 
on the ionic side \cite{soosklein} in many segregated stacks
 and in conjugated polymers \cite{handbook,sooshay}.
Theoretical interest in $H_0(t,\Delta ,U)$ and its variants 
lies in the interplay of e-e and e-ph interactions and their 
role in the structural instabilities of 1-D materials, 
broadly defined \cite{torino,handbook,sooshay,nasu}.

The $t = 0$ GS of the Hamiltonian in Eq.~(\ref{uno})
 is sketched in Fig. 1 to illustrate some basic features. 
The electrons are paired on D (odd) sites for $\Delta > U/2$, 
paired on A (even) sites for $\Delta < -U/2$, 
and singly occupy all sites in between; covalent states have 
$n_i = 1$ at all $i$ and spin degeneracy of $2^N$. 
The valence transitions at $U = \pm  2\Delta$  for $U > 0$
 transfer one electron in the GS and $\rho = 2 - n_D$
 changes discontinuously. Since the paired GS in Fig. 1 are nondegenerate, 
we expect finite gaps for spin, optical and charge carrying excitations. 
The covalent GS, on the other hand, has vanishing spin gaps.

$H_0(t,\Delta,U)$ with finite $t$ is a one-dimensional metal at
$\Delta  = U = 0$, a band insulator for $\Delta > 0$, $U = 0$, 
and a Hubbard model for $\Delta= 0$, $U > 0$. 
Finite $t$ leads to continuous $n_D$ at the NIT when $\Delta$
 is not a function of $n_D$. Although the ionicity is continuous, 
the NIT between band and Mott insulators is a true quantum 
phase transition at T=0K as signaled by the closing of 
triplet \cite{sbm} and singlet \cite{gp86,pg88} gaps 
and by the unconditional instability \cite{gp86,pg88}  to dimerization 
on the covalent side. The nature of the transition between 
two insulators has revived interest in the modified Hubbard 
model in connection with localization and conductivity 
in correlated systems \cite{resta}.
 Strong charge fluctuations induced by lattice 
motion \cite{egami,resta} and related structural instabilities \cite{fabrizio}
 have been rediscovered and underlined. Finite $t$
 generates correlated states of the Hamiltonian in Eq.~(\ref{uno})
 that differ fundamentally from the $t = 0 $ limit. 
In spite of sustained research 
\cite{egami,soosmazu,gp86,pg88,pg92,nagaosa,avignon,resta,fabrizio,ortiz,tosatti,caprara,takada,qin}, no definitive picture has emerged for the T=0K 
phase diagram of the simple model in Eq.~(\ref{uno}).  

We present in this paper exact solutions of $H_0(t,\Delta ,U)$ for finite 
$N$ using valence bond (VB) methods \cite{vb} that were originally 
developed \cite{soosmazu} for CT complexes. 
Total spin $S$ is conserved in all versions of Eq.~(\ref{uno}). 
VB diagrams with specified pairing of sites with $n_i = 1$
 form a large but complete basis for any $N$. 
The scope of finite-$N$ results is decisively extended 
by using both periodic and antiperiodic boundary conditions 
and the symmetries of $H_0$. 
The oligomers in Section 2 reach $N$ = 16 in the full basis 
or $N$ = 22 in a restricted basis without D$^{2+}$ or A$^{2-}$ sites, 
compared to $N \sim$ 10 in previous studies.
 Exact excitations near the NIT are related in Section 3 to the 
opening of gaps and interpreted as due to localization on the paired side.
 The GS is metallic at the NIT according to the charge gap and 
charge stiffness. 
The NIT  marks the boundary between a localized GS 
 for $\Delta >> U$, and a delocalized GS 
with vanishing excitation energies as in Hubbard models 
at $\Delta= 0$. We shall briefly mention the role of e-ph 
coupling and intersite e-e interactions, but defer detailed 
analysis to subsequent publications.

\section{ Symmetry crossover and charge density }

We consider GS properties of $H_0(t,\Delta,U)$
 at half filling, one electron per site, and uniform $t $ = 1. 
General solutions of the Hamiltonian in Eq.~(\ref{uno})
 are restricted to finite $N$, where eigenstates and 
 energies are accessible. Periodic boundary conditions 
(PBC) are readily applied to noninteracting ($U =0$) systems whose GS
 energy is

\begin{equation}
\frac{E_0(\Delta , 0)}{N} =-\frac{1}{N}
\sum_{k ~ filled} 2(\Delta ^2 +4 \cos^2 {k} )^{1/2}
\rightarrow -\frac{2}{\pi}(\Delta ^2 +4 )^{1/2}  E(q)
\label{tre}
\end{equation}
The expression for the infinite chain is shown in Fig. 1; 
$E(q)$ is the complete elliptic integral of the second kind, 
with $q^2 = 4/(\Delta ^2+4)$. From  Eq.~(\ref{due})
the  GS electronic density on D is:

\begin{equation}
n_D(\Delta ,0) -1 = \frac{2\Delta K(q)}{\pi (4+\Delta ^2)^{1/2}}
\label{quattro}
\end{equation}
where $K$ is the complete elliptic integral of the first kind. 
The divergence of $\partial n_D/\partial \Delta $ at $\Delta = 0$
 signals an electronic instability. The behavior of finite rings
 is different and shows 4n, 4n+2 effects. The wavevector is 
$k = 0, \pm 2\pi /N, \pm 4\pi /N, ...\pi $. We have energies
$\pm \Delta$ at $k = \pi /2$ when $N$ = 4n and two electrons for 
these orbitals. The degeneracy produces an energy cusp at $\Delta = 0$;
 $n_D$ changes discontinuously and the partial derivative
in Eq.~(\ref{due}) is not defined. Finite rings with $N$ = 4n+2 
have nondegenerate GS at $\Delta = 0$, no cusp and finite 
 $(\partial n_D/\partial \Delta)_0 $.
  The 4n, 4n+2 sequences must coincide in the extended chain 
and do so according to Eq.~(\ref{quattro}), with continuous
 $n_D$ and divergent  $(\partial n_D/\partial \Delta)_0 $. 
Exact $U = 0$ results illustrate the extrapolation problems 
encountered in interacting chains.

The full basis of $H_0(t,\Delta,U)$ increases roughly as $4^N$ 
with $N$ and as $3^N$ when we exclude doubly ionized sites, 
i.e. two electrons 
 at A sites or two holes  at D sites. 
We use VB methods \cite{vb} to reach $N$ = 16 
for the full basis and $N$ = 22 for the restricted basis. 
The basis has over $10^7$ singlets or $10^9$ Slater determinants with 
$S_z = 0$. Exact solution \cite{lieb,ovchi}
 of the extended chain is limited to $\Delta = 0$, the Hubbard model. 
The GS is a nondegenerate singlet, the charge gap is finite for $U > 0$
 and there is spin-charge separation at large $U$. 
Finite $t$ and $U$ always lowers the energy in Fig. 1 compared to 
$t = 0$. The greatest changes occur at $\Delta = \pm U/2$, 
where $t$ cannot be treated as a small parameter.

In $C_N$ symmetry, the GS of the interacting systems transforms as 
$k' = \pi$ on the covalent side of 4n rings and as $k' = 0$ in 4n+2 rings. 
Site energies $\Delta > 0$ lower the symmetry from $C_N$ to $C_{N/2}$
 and yield a charge-density-wave (CDW) GS. 
The extended system no longer has inversion centers between sites, 
which corresponds to reflection between sites for finite $N$, 
but retains inversion at the sites or, for finite $N$, reflection 
$\sigma_v$ through the sites. 
With two sites per unit cell, both $k' = 0$ and $\pi$ transform as 
$k = 0$ in the first Brillouin zone. 
According to reflection through sites, the covalent GS of 4n+2 rings 
is even ($\sigma_v = 1$, $A_1$) and that of 4n rings is odd 
($\sigma_v = -1$, $A_2$). The GS of 4n rings is degenerate at 
$\Delta_c(U,N)$, where the symmetry switches from 
$A_2$ to $A_1$ with increasing $\Delta$, and this crossover 
defines the NIT. In addition to PBC, we use antiperiodic boundary 
conditions (APBC) with reversed sign $t_{1N} = -1$ 
for transfer between 1 and $N$.
This corresponds to $\phi _{n+N}=- \phi _n $
for the on-site wavefunctions and a periodicity 2$N$.
In terms of VB diagrams, we modify $\sigma_v$ for 
reflection through sites 1 and $N/2 +1$ to:

\begin{equation}
\sigma ' = \sigma _v (-1)^{n_1}
\label{cinque}
\end{equation} 
where $n_1$ is the occupation number of site 1. 
The APBC GS has $\sigma ' = 1$ in 4n rings for any $\Delta , U$. 
The GS of 4n+2 rings have a crossover from $\sigma ' = -1$ at small 
$\Delta $ to $\sigma ' = 1$ at $\Delta >\Delta  _c(U,N)$. 
The subspaces $A_1 ', A_2'$ associated with $\sigma '$
 do not coincide with $A_1$, $A_2$. The paired state is 
unique and even for either PBC or APBC. There are two covalent states, 
the Kekul\'{e} diagrams for benzene, with nearest-neighbor pairing 
of all spins. We define $|K1\rangle$ and  $|K2\rangle$ as pairing
spins at sites $2i-1$, $2i$ and $2i$, $2i+1$, respectively, for all $i$. 
The pairing in $|K1\rangle$  is D$^+$A$^-$, while the pairing in 
 $|K2\rangle$ is A$^-$ D$^+$. The combination $|K1\rangle +|K2\rangle$  
 transforms as $A_1$ or $A_2'$ for PBC and APBC, respectively, 
while the out-of-phase combination transforms as $A_2$ or $A_1'$.

The $U = \Delta = 0$ crossover connects electrons paired as 
D$^{+2}$A$^{-2}$ or DA in Fig. 1. For $U > 0$, the NIT shifts to 
positive $U - 2\Delta _c$ and $t\neq  0$ preferentially stabilizes
 the paired GS over the covalent GS because the latter 
has finite probability for adjacent parallel spins that cannot
 transfer under Eq.~(\ref{uno}). The symmetry changes at 
$\pm \Delta _c(U,N)$ in rings with either PBC or APBC. 
Exact crossovers are shown in Fig. 2 as $U - 2\Delta _c(U,N)$ in the 
$U,\Delta >0$ quadrant for $U$ = 0.5, 1, 2, 3, 4, 5, and 10; 
the inset has $U$ = 100, 200, 300 and $\infty$, the last one corresponding to
the restricted basis. At fixed $U$ and finite $N$, 
the crossovers are similar for 4n with PBC and 4n+2 with APBC. 
The dashed line is an extrapolation to the infinite chain discussed below. 
The covalent region  
is very narrow and the crossovers merge at $U = 0$. 
The inset shows that $\Delta _c(U,N)$ is nearly constant for
$\Delta  > 5t$.

We plot $\Delta _c(U,N)$ vs. $N^{-2}$ in Fig. 3 and find accurate 
extrapolation at large $\Delta , U$. The difference between 
$U$ = 300 and the restricted basis is due to small admixtures of 
A$^{2-}$D$^{2+}$ at energy $ U+ 2\Delta$. The extrapolated limit is 
$U - 2\Delta _c =$ 1.332 in the restricted basis.  
It has previously been estimated
\cite{gp86} as 1.2-1.3 based on the singlet and 
triplet gaps, respectively, of $N \leq 10$ rings and \cite{sbm} at 
1.5 based on the ionicity up to $N = 10$. Mixing with A$^{2-}$D$^{2+}$
 grows as $U$ decreases, as seen for $U = 10$, and the functional 
dependence is closer to $\sim  N^{-1}$ at smaller $U \sim 3$. 
Kinetic contributions are largest at small 
$U$ and $\Delta$, where $t$ is comparable to CT energies. 
The extrapolated (dashed) line in Fig. 2 is based on a power law,
$\Delta _c(U,N) \propto N^{-\gamma}$, with $1 < \gamma < 2$
 giving the best fit for each $U$ from $N$ = 8 to 16.

Figures 2 and 3 indicate that, except for $\Delta, U < 2t$, 
the NIT hardly varies with $U/\Delta$. The relevant 
DA systems have narrow bands,
oxides are modeled \cite{egami,resta}
 with wider bands, $\Delta < t$. The restricted basis captures 
the basic physics. We let both $\Delta$ and $U$ diverge in 
$H_0(t,\Delta,U)$ while keeping $\Gamma= \Delta - U/2$ finite 
\cite{soosmazu} and reference the crossover to $\Gamma - \Gamma_c =
\Delta  -\Delta _c$.  In the half-filled case, $\Delta \rightarrow \infty$
 ensures an electron at each D site and excludes two at any A site. 
The $t = 0$ GS has energy $-2\Gamma$, 0  per DA on the paired and 
covalent side, respectively, with $\Gamma = 0$ at the NIT. 
The restricted basis is almost quantitative for $U > 5$, makes
larger $N$ accessible, and holds at the NIT. The related limit with both $U$
and $U-2\Delta >> t$ leads instead to a Heisenberg spin chain 
\cite{nagaosa} without charge degrees of freedom and does not apply to 
the NIT.

The GS expectation value, $\langle n_{2i-1}\rangle$, for electrons at 
D sites is more accurate than the numerical derivative in Eq.~(\ref{due}). 
Matrix elements \cite{vb} over correlated states can be evaluated 
exactly for finite $N$. Results for the restricted basis of 4n rings with 
PBC and 4n+2 rings with APBC are shown as a function of 
$\Gamma  - \Gamma _c(N)$ in Fig. 4a. The crossover generates a jump in 
$n_D$. The charge density is continuous in 
$A_1$, $A_2$ for 4n rings, or in $A_1'$, $A_2'$ for 4n+2 rings, 
but continuing the lines in Fig. 4a through the NIT gives an 
excited-state density. All approaches to the NIT described by the hamiltonian
in Eq.~(\ref{uno}) indicate $n_D$ to be continuous when $t$ is finite.
As expected, the discontinuity in $n_D$ decreases with $N$
 and vanishes in the extended chain. The GS of 4n+2 rings with 
PBC or 4n rings with APBC remains in $A_1$ or $A_1'$, respectively, 
for any $U, \Delta$ and the charge density is continuous, as shown in
 Fig. 4b. The NIT defined by the maximum of $\partial n_D/\partial \Delta$
 is less precise numerically (by $\sim 0.02 t$) than a crossover. 
The curves in Fig. 4b have been adjusted to catch the extrapolation 
between increasing and decreasing series on either side of $\Delta_c$. 
Results for the infinite chain are shown as stars that coincide 
in both panels. They represent joint extrapolations as either $N^{-1}$ or 
$N^{-2}$ that give the smaller mean square deviation. 
Indeed, $n_D$ is almost quantitatively known from the 
requirements that $n_D(22) > n_D(20)$ on the covalent side and 
$n_D(22) < n_D(20)$ on the paired side. The present estimate for $n_D$
is 1.314(2) at the NIT, i.e. $\rho = 0.684$. 

The restricted basis for $U, \Delta >> t$ fixes $n_D = 1.31$ at the NIT of
Eq.~ (\ref{uno}). The slope $\partial n_D/\partial \Delta$
  is finite, but this is inconclusive by itself. 
H\"uckel rings show similar 4n, 4n+2 behavior and exact $N$ = 200 and 400 
results are indistinguishable from  $n_D$
 in Eq.~(\ref{quattro}) at the resolution of Fig.~(4). The origin must be 
magnified an order of magnitude to see the divergence of 
 $\partial n_D/\partial \Delta =  (\partial ^2 E_0/\partial \Delta^2)/N$
 at $\Delta = 0$. This divergence signals the intrinsic instability of 
the $U= \Delta = 0$ chain to a site-CDW distortion that, however, 
is already broken at finite  $U = 2\Delta _c$ in interacting systems. 
Extended chains with $U>0$ have finite  $\partial n_D/\partial \Delta $
at the NIT. By contrast, the Peierls instability to a bond-CDW
is unconditional \cite{pg88,nagaosa} for any $t/U$, because dimerization 
breaks reflection symmetry  $\sigma_v$,  as experimentally 
recognized in the initial 
TTF-CA studies \cite{girlando}. 

The full basis of the Hamiltonian in Eq.~(\ref{uno}) is required for 
small $U$ and exact results for $n_D$ or $\Delta _c(U)$
are limited to $N$ = 16. We again have discontinuous $n_D(\Delta,U)$
 in 4n rings with PBC and 4n+2 rings with APBC, and continuous $n_D$
 for the opposite boundary conditions. We compare in Fig. 5 extrapolated
$ n_D$ for $U =$ 2, 5 and 10 with the exact $U = 0$ result 
in Eq.~(\ref{quattro}). The arrows marking the NIT for finite $U$
 are based on symmetry crossovers and extrapolations similar 
to Fig. 3. We have increasing $n_D(\Delta_c,U)$ with $U$ and the limiting 
value of $\sim $1.3 is reached by $U = 5$. Figure 5 
shows the stabilization of covalent states with increasing 
$U$ and small NIT variations for $\Delta , U > 2$.

\section{  Energy gaps and localization at the NIT }

The excitations of $H_0(t,\Delta ,U)$ provide other evidences of the NIT. 
The paired and covalent states are diamagnetic and paramagnetic, 
respectively. A singlet-triplet gap, $E_{ST}$, opens \cite{sbm} at the NIT 
between a band insulator with $E_{ST} > 0$ and a Mott insulator 
with $E_{ST} = 0$. The lowest singlet excitation, $E_{SS}$, 
is between the $A_1$ and $A_2$ GS; hence $E_{SS}$ vanishes \cite{gp86}
 at the NIT. The transition is dipole allowed and is formally the 
CT excitation, but $E_{SS}$ rapidly loses oscillator strength on the 
covalent side. The charge degeneracy in Fig. 1 at $ U = 2\Delta$, $t = 0$
 distinguishes between neutral and ionic complexes. 
The $t > 0$ gaps near the NIT are 
not known and their simultaneous opening, as tacitly supposed for a 
single transition,\cite{soosmazu,gp86,pg88,pg92,nagaosa,avignon,horovitz}
 is not assured nor agreed on 
\cite{resta,fabrizio,ortiz,tosatti,caprara,takada,qin}. 
We report exact excitation thresholds
 near the NIT defined by GS crossovers. All symmetry considerations 
apply to the full basis for $\Delta > 0$.

Figure 6 reports $E_{SS}(N) = E_2(N) - E_1(N)$, i.e. the energy difference 
between $A_2$ and $A_1$ GS, in the restricted basis as a function 
of $\Gamma -\Gamma _c(N)$. Since $E_{SS}(N)$ increases in 4n rings 
for $\Gamma > \Gamma _c$ and decreases in 4n+2 rings, we have finite 
$E_{SS}$ on the paired side. On the covalent side, $E_{SS}(N)$
 decreases with $N$ in rings whose GS remains in $A_1$ or $A_1'$ 
and increases in rings whose GS is in $A_2$ or $A_2'$. 
Joint extrapolations yield the stars that are consistent with vanishing 
$E_{SS}$ in the extended system. Exact results to $N$ = 22 in Fig. 6 
are the most stringent limit to date, with $E_{SS} < 0.05 t$
 on the covalent side and finite $E_{SS}(N)$  for 
$\Gamma - \Gamma_c < 0.05t$. We note that at $\Delta = 0$, 
far on the covalent side, Ovchinnikov \cite{ovchi}
 found nonpolar singlets with zero gap for any $U > 0$. 
Far on the paired side, we have $E_{SS} \sim  2\Gamma$
 by inspection. Hence increasing $\Delta$ at fixed $U$ in 
the extended system clearly opens an SS gap that is seen to coincide in 
Fig. 6 with the NIT defined by the symmetry crossover. 
The unconditional instability for dimerization on the covalent side 
is closely related to vanishing $E_{SS}$; the instability is 
conditional for a finite gap. 

The magnetic gap $E_{ST}$  is to the lowest triplet for either 
PBC or APBC. As shown in Fig. 7, $E_{ST}$ increases rapidly with 
$\Gamma >\Gamma _c$ in the restricted basis and is small 
on the covalent side. Open circles represent boundary conditions 
with crossovers and systems whose $E_{ST}$  increases with $N$ at larger
$ \Gamma - \Gamma _c$. Closed circles are for boundary conditions 
without crossovers and show decreasing $E_{ST}$ with $N$. 
The stars in Fig. 7 are joint extrapolations. The bound on $E_{ST}$
 is $E_{ST} < 0.1t$ for $\Gamma  < \Gamma _c$ and the gap opens 
at $\Gamma _c$ or slightly on the covalent side. At finite $U$, 
the extended system is rigorously known to be paramagnetic \cite{taka}
 at $\Delta = 0$, with $E_{ST} = 0$, and diamagnetic with 
$E_{ST} \sim 2\Delta - U$  for $\Delta >> U$. 
The opening of an ST gap with increasing $\Delta$ is assured, and 
the results in Fig. 7 are consistent with $E_{ST} > 0$ at the NIT. 
The concomitant dimerization on the covalent side opens a magnetic gap, 
as it is well known in spin chains \cite{soosklein} with
regular or alternating exchanges and triplet spin excitons.

The charge gap of the Hamiltonian in Eq.~(\ref{uno}) 
is $I-A$, since there is not an explicit Madelung contribution. 
$I-A$ is related to  the GS of the cation and 
anion radicals, $E_+(N)$ and $E_-(N)$, respectively, and corresponds 
to charge disproportionation or electron transfer between 
noninteracting systems,

\begin{equation}
I(N) -A(N) = E_+(N) +E_-(N) -2E_0(N).
\label{sei}
\end{equation}
At $t = 0$ and $\Delta  > 0$, we have a paired GS for $U < 2\Delta$ 
with $I = \Delta - U$ and $A = -\Delta$; the lines cross at the NIT 
and the covalent side has $I = -\Delta$ and $A = \Delta - U$ for 
$U > 2\Delta$. For $t > 0$, the charge gap of free electrons, 
$|2\Delta |$, follows from the valence and conduction bands in 
Eq.~(\ref{tre}); the extended $U = 0$ system is metallic at
 $\Delta = 0$ and insulating otherwise. Although not known exactly 
for $U > 0$, the charge gap is readily shown to be large, 
roughly $|2\Delta -U|$, far from the NIT. On the covalent side, 
it becomes the Lieb-Wu gap \cite{lieb} at $\Delta = 0$
 and increases as $U$ for $U > 4t$; on the paired side, 
all gaps increase as $2\Delta - U$ for $\Delta >> U$. 
Finite $N$ leads to charge gaps at the NIT in systems with discrete energies.

Both e-h symmetry \cite{mclac} and $\sigma_b$ are broken for $\Delta > 0$, 
but their product remains a symmetry operation. Extended e-h symmetry
\cite{skh} cuts basis, either full or restricted, roughly in half 
and corresponds in the $S = 0$ manifold to the $A_g^+ \oplus  B_u^-$
 and $A_g^- \oplus  B_u^+$ subspaces of $H_0(t,0,U)$. 
The GS symmetry does not change at the NIT. E-h symmetry relates the 
GS and excited states of the radical ions \cite{mclac,skh}.
 In particular, we have $E_-(N) = E_+(N) + U$, a general result 
that holds on adding any spin-independent potential to $H_0(t,\Delta,U)$. 
It follows that Eq.~(\ref{sei}) reduces to $I(N) + A(N) = -U$
 for even $N$ and arbitrary $t, \Delta , U$. 
Table 1 reports $I(N) - A(N)$ at $U, \Delta _c$ up to $N$ = 14 
in the full basis and $N$ = 18 in the restricted basis. 
The $U = 0$ gaps vanish at the crossover, where the electron transfer
described in Eq.~(\ref{sei}) involves degenerate orbitals. 
The charge gaps increase with $U$ but remain small at the crossover 
even for divergent $U$. The gaps in Table 1 follow power laws, 
$N^{-\gamma}$, with $\gamma < 0.6$ and place a rough bound of $\sim 0.2t$
 on the extended system. The charge gap vanishes at most at a single point,  
$\Delta_c(U)$ or $U_c(\Delta)$, that coincides with the NIT within 
the accuracy of finite systems.

The SS, ST and charge gaps are all finite on the paired side, 
$\Gamma >\Gamma _c$. They differ on the covalent side, 
however, where only the charge gap is finite. We associate gaps 
on the paired side with localization. The GS for 
the $|\Delta | >> U$ limit has paired spins on either odd or 
even sites and is manifestly localized. Since gapless triplets 
and singlets are firmly established \cite{ovchi} at $\Delta = 0$, 
the GS of $H_e(t,\Delta,U)$ for $U, t > 0$ is extended at $\Delta = 0$ 
and localized at $\Delta >> U$. A localization-delocalization 
transition between two insulators incorporates all aspects of the NIT 
and the vanishing charge gap suggests a metal at the transition. 
We develop these ideas below. 

To show localization on the covalent side, we partition $H_0$
 in the restricted basis into $h_0$ for transfers between sites 2n-1 
and 2n, as in $|K1\rangle$,  and a perturbation $V$ for transfers between 
2n and 2n+1. We take $\Gamma=  \Delta - U/2 > 0$ in units of $t$
 and solve the 2x2 dimer problem in the singlet subspaces of $h_0$. 
The exact GS of $h_0$ is

\begin{equation}
|G_0(\Gamma )\rangle = \prod _{i=1}^{N/2} \left[ 
a^+_{2i-1,\alpha}a^+_{2i-1,\beta}
\cos{\phi} +\sqrt{2} \left(  a^+_{2i-1,\alpha}a^+_{2i,\beta}-
 a^+_{2i-1,\beta}a^+_{2i-1,\alpha}\right) \sin{\phi}\right] |0\rangle
\label{sette}
\end{equation}
where $|0\rangle$ is the vacuum state and $\tan{2\phi} = \sqrt{2}/\Gamma$ 
 governs the mixing of $|DA\rangle$ and the singlet linear combination of 
$|D^+A^-\rangle$. The zeroth-order energy is
 $-\Gamma -(\Gamma^2 +2)^{1/2}$  per dimer. 
The opposite choice of 2n, 2n+1 for dimers has the same energy but 
admixes $|A^-D^+\rangle$ singlets, as in $|K2\rangle$. 
Each dimer has a triplet with excitation energy 
$\epsilon_T = \Gamma + (\Gamma^2 +2)^{1/2}$, a singlet at 
$\epsilon_S = 2(\Gamma^2 +2)^{1/2} $ and strictly confined electrons.  
The perturbation 

\begin{equation}
V=-\sum_{i,\sigma} \left( a^+_{2i,\sigma} a_{2i+1,\sigma}+
  a^+_{2i+1,\sigma} a_{2i,\sigma} \right),
\label{otto}
\end{equation}
is necessarily small when $ \Gamma $ is large. 
To second-order in $V$, the energy per dimer is

\begin{equation}
\epsilon^{(0)} +\epsilon^{(2)} = -\Gamma -(\Gamma^2+2)^{1/2} -
\frac{\cos^4{\phi}+\frac{\sin^4{\phi}}{4}}{(\Gamma^2 +2)^{1/2}}
\label{nove}
\end{equation}
Electrons are now confined to adjacent dimers that are connected 
by virtual excitations. As shown in Table 2, 
Eq.~(\ref{nove}) is nearly quantitative as close to the NIT as $\Gamma = 2$. 
Localization to adjacent dimers approximates the exact solution of 
Eq.~(\ref{uno}), which for $N$ = 22 is a linear combinations of over 
$10^7$ singlets. Rapid convergence with $N$ also points to localization 
on the paired side and is seen for $n_D$ in Fig. 4,
$ E_{SS}$ in Fig. 6 and $E_{ST}$ in Fig. 7. Successive orders in $V$
 increase by one the number of coupled dimers. 
Such an expansion fails at the NIT or on the covalent side. 

To see if the system is metallic at the transition, we compute the 
charge stiffness \cite{kohn} relevant to the Hamiltonian in Eq.~(\ref{uno}). 
This property has been applied to interacting fermions
\cite{ss,sap,tsiper} in one dimension. The perturbation is a phase 
factor $\exp{(\pm if)}$ in Eq.~(\ref{uno}) for transfers to the right and 
left \cite{sap},

\begin{equation}
V(f) = (1-\cos{f})\nu_+ +i\nu_- \sin{f}
\label{dieci}
\end{equation}
The first term of Eq.~(\ref{uno}) is $-\nu _+$, while $\nu _- $ 
has oppositely signed transfers to the right and left and connects 
$A_1$ and $A_2$ states for PBC. We now have $E_0(\Gamma,t,f)$ in units of $t$. 
The charge stiffness per site is 
$\chi _{cs} =(\partial ^2 E_0 / \partial f^2)_0/N$. 
It is finite in conductors and vanishes in insulators. 
At the crossover, the proper zeroth-order GS of 4n rings is the odd 
linear combination of the $A_1$ and $A_2$ GS and

\begin{equation}
\chi_{cs}(\Gamma _c) = \left( \frac{\partial^2 E_0}{N\partial f^2} \right)
_0 = p^{(1)} +p^{(2)}
= \frac{|E_0|}{N} +\frac{\Gamma _c (n_D^{(1)} +n_D^{(2)}-2)}{2}
\label{undici}
\end{equation}
Here $n_D^{(1)}$ and  $n_D^{(2)}$ are the electron densities at donor sites 
 in the $ A_1$ and $A_2$ subspaces, respectively, and 
 $ p^{(1)}$ and $ p^{(2)}$  are the corresponding bond orders, with 
  $2p$ defined as the  GS expectation value of $\nu_+$, 
which we evaluate in the restricted basis. 
The donor densities at the NIT are shown in Fig. 4 and have opposite
 $N$ dependence in $A_1$ and $A_2$. The value of $\Gamma_c = U/2 - \Delta_c 
 = -0.666$ follows from Fig. 3. 
We have a poor metal: $\chi_{cs}(\Gamma _c) \sim 0.74$ is $\sim$60\% of 
$4/\pi$  value for free electrons at $\Delta = U = 0$ in 
Eq.~(\ref{uno}). For $\Gamma \neq \Gamma_c$, 
second-order perturbation theory in $f$ becomes exact
\cite{sap}. Fig. 8 shows that  $\chi _{cs}(\Gamma)$ 
is exponentially small on either side of the NIT. 
The charge stiffness \cite{ss} of the Hubbard model 
has a similar peak at $U=0$ that narrows with $N$ and 
becomes a $\delta$-function in the infinite chain.
Hence we expect  $\chi_{cs}(\Gamma)$ to vanish 
except at $\Gamma_c$ in the extended system.

A metallic point connecting insulating phases has been recently  
discussed for $H_0(t,\Delta,U)$ \cite{resta} and for 
the following half-filled system of spinless fermions \cite{tsiper},

\begin{equation}
H= -\sum_i (a^+_{i}a_{i+1} +a^+_{i+1}a_{i}) + 
   \sum_i ({\mathcal V} n_i n_{i+1} +{\mathcal W}n_i n_{i+2})
\label{dodici}
\end{equation}
Large ${\mathcal V}> 0$ favors a GS without adjacent occupied sites, 
while large ${\mathcal W}$ 
favors one with adjacent filled and empty sites along 
the chain. The $t = 0$ crossover occurs at ${\mathcal V} = 2{\mathcal W}$, 
with adjacent 
electrons and holes for $ {\mathcal V}< 2{\mathcal W}$ that resemble 
D and A sites, respectively. 
The ${\mathcal V} >2{\mathcal W}$ GS has alternating filled and 
empty sites that, taken in pairs, 
correspond to D$^+$ and A$^-$. 
The crossover is not precisely at ${\mathcal V} = 2{\mathcal W}$, 
presumably due to 
different bond orders in the two GS. It shifts to 
$ {\mathcal W} - {\mathcal V}/2 \sim -0.6$
 in units of $t$, very close to $\Gamma_c$. Transfers between 
two sites in Eq.~(\ref{dodici}) differ from spin degeneracy in 
 Eq.~(\ref{uno}), however, and the models do not map into each other. 
Exact results \cite{tsiper}
 to $N $ = 40 for Eq.~(\ref{dodici}) are comparable to 
$N$ = 20 for the full basis of Eq.~(\ref{uno}). 
The striking similarity between Fig. 8 and the charge stiffness of 
 Eq.~(\ref{dodici})  up to $N$ = 40 suggests a similar interpretation.

\section{discussion}

The modified Hubbard model 
in Eq.~(\ref{uno}) has many applications to both theory and experiment.
 With variations, it is suitable for modeling valence transitions, 
excitation thresholds, electronic or structural instabilities,
 among other topics. Its two parameters, $U/t$ and
 $\Delta/t$, encompass the Hubbard model ($\Delta = 0$) 
at half or other filling, two bands at $U = 0$ and localized dimers
for $\Delta >> U$. At fixed $U$ and $t$, increasing $\Delta > 0$
 generates a neutral-ionic transition whose characterization 
is the principal goal of this paper. The NIT of the Hamiltonian
in Eq.~(\ref{uno}) has continuous ionicity given by 
Eq.~(\ref{due}) and excitation gaps for singlets, 
triplets and charges that are not known exactly. 
Previous approximations have been developed separately for 
$n_D$, $E_{SS}$, $E_{ST}$, the charge gap, the GS at the NIT, 
instabilities, etc. Our collective analysis 
of symmetry crossovers, excitation thresholds and GS properties
incorporates   computational advances and yields better estimates 
for extended systems.

Finite-size results require extrapolations whose accuracy improves with $N$. 
We followed the NIT of $H_0(t,\Delta,U)$
 through the symmetry crossover of the GS, the charge density 
$n_D$, the excitations $E_{SS}$, $E_{ST}$, and the charge gap 
and stiffness. Larger $N$ is accessible in the restricted basis 
allowing for an accurate estimate of $\Gamma _c =-0.666t $
from the crossover and setting 
stringent limits of $\sim  0.1 t$   for the opening of all three gaps
at this position. Thus the numerical results point to a single
 transition. 
The NIT of the modified Hubbard model is continuous, as previously 
found, and is marked by the opening of singlet, 
triplet and charge gap on the paired side. There is no gap in the singlet 
or triplet manifold on the covalent side. 
The interacting system is known to have a delocalized GS at 
$\Delta = 0$, the Hubbard limit, and localized GS for $\Delta >> U$,
 the paired limit. We identify the NIT at 
$\pm \Delta _c(U,t)$ as the appearance of a localized GS.

Resta and Sorella discuss \cite{resta} polarization and 
metallic behavior at the NIT in the context of oxides, 
with $t_0$ = 3.5 eV, $\Delta '$ = 2.0 eV and variable $U$ in 
Eq.~(\ref{uno}). The crossover in $N$ = 8 rings, at $U/t_0 = 2.27$, 
is used to estimate the polarization of extended systems. 
Since $\Delta '/t_0$ = 0.571 corresponds to large $t$, the crossover 
is near the origin of the $\Delta , U$ plane in Fig. 2 and there are 
substantial finite-size effects. The $N$ = 8 result in Fig. 2 yields 
$U_c = 2.27$, in quantitative agreement with ref. \cite{resta}, 
but larger $N$ up to 16  extrapolate to larger $U_c/t$ = 2.70 
for the extended system. Such  corrections are consistent with 
 $N \sim 10$ results on CT complexes. The GS polarizability 
is a new approach, different from the charge stiffness, to the 
identification of metallic behavior.

The GS density is $n_D$ = 1.314 at the NIT of the restricted basis, 
when one electron is always confined to D. 
The spin degeneracy of D$^+$ or A$^-$ spoils exact analysis. 
The degeneracy of charge and spin excitations at the NIT gives 
a simple, heuristic interpretation: $n_D = 4/3$
 is the result for equal weights of molecules and spin-1/2 
radical ions. Equal weights at the NIT can be justified rigorously at 
$U = 0$ for electrons or for spinless fermions, but not in the restricted 
basis. We found $\Gamma _c$ = -0.666 in the restricted basis and use this 
value in $|G_0(\Gamma)\rangle$, the dimer GS in Eq.~(\ref{sette}); 
the paired-state amplitude is $\cos ^2{\phi} $= 0.287, 
which corresponds to $n_D$ = 1.287. Dimers capture most of the 
configuration mixing of the extended system. 
The full basis has contributions from D$^{2+}$ and A$^{2-}$ diagrams, 
which as seen in Fig. 5 reduce $n_D$ compared to the restricted basis.

Peierls-Hubbard models are widely applied to structural instabilities. 
The stability of the GS to a perturbation can be formulated 
in terms of susceptibilities, $\chi$, that are formally given by the 
exact eigenstates $|F\rangle$ and energies $E_F$ of the
hamiltonian in Eq.~(\ref{uno}). The perturbation is written as the
product 
$\theta Q$, where $\theta$ is the relevant operator for coupling to 
$Q$, and the corresponding $\chi$ is

\begin{equation}
\chi \propto - \left( \frac{\partial ^2 E_G} {\partial Q^2}\right) _0
 = 2\sum_F \frac{|\langle G|\theta |F\rangle |^2}{E_F-E_G}
\label{tredici}
\end{equation}
Since the sum is over the excited states of the unperturbed system, 
the eigenstates of the uniform chain in Eq.~(\ref{uno}) suffice
 for the stability of the modified Hubbard model. 
The charge stiffness in Eq.~(\ref{dodici}) is $\chi$ with 
respect to a magnetic field perpendicular to the ring \cite{sap}
 and gives information about current flow.
Structural transitions are investigated by introducing phonons as 
$Q$-perturbation. The Peierls instability for dimerization involves 
$k = 0$ phonons with $\theta $ representing the staggered bond-order 
operator (the first term in Eq.~(\ref{uno}), 
augmented by a $(-1)^i$ 
factor). This operator breaks inversion symmetry at the sites and mixes 
$A_1$ and $A_2$ singlets \cite{gp86}. Vanishing $E_{SS}$ on the covalent side 
of the NIT then indicates a divergent $\chi$ and the unconditional 
instability of a lattice with harmonic potentials \cite{pg88}. 
On-site (Holstein) phonons couple instead to CDW operator $n_D$. 
Since $\chi_\Delta = \partial n_D/\partial \Delta$
 is finite at the NIT, except for $\Delta = 0$, 
the corresponding instability is conditional \cite{pg88}; 
the NIT marks the maximum  $\chi_\Delta$, i.e. the maximum 
$\partial  n_D/\partial \Delta$, 
as discussed under Fig. 4.

We turn next to open or controversial aspects of the NIT of the 
modified Hubbard model. Some authors \cite{fabrizio,takada}
 have proposed two transitions related to the closing of charge and spin gaps,
 respectively; a spontaneously dimerized phase then separates a band 
insulator corresponding to the paired GS and Mott insulator 
on the covalent side \cite{fabrizio}. 
The suggestions \cite{fabrizio}
for another transition rest on the analogy with spin-1/2 Heisenberg
antiferromagnetic chains with frustration due to a second-neighbor exchange
$J_2$. The Kekule diagram $|K1\rangle$  or $|K2\rangle$
 is the exact GS at $J_2 = J_1/2$, as
recognized by Majumdar \cite{majumdar}. 
There is no exact mapping of Eq. (1) into
such a spin chain, not even at large $U$, but the GS of related models with,
for example, second-neighbor transfers have not been studied in detail.
Our exact results for $H_0(t,\Delta, U)$ with finite $N$ show that 
the maximum of $\partial \rho/\partial \Delta$, the closing of 
the singlet and triplet gaps, and the vanishing of the charge gap
coincide at the NIT  within  $\sim 0.1 t$. Finite systems cannot
specify transitions, but provide some constraints. 
The Mott insulator dimerizes spontaneously 
for any $t/U$,
 as discussed for the spin-Peierls instability
\cite{bray}
of Heisenberg antiferromagnetic chains. The dimerization amplitude
 becomes very small for $U >> t$ and $J = t^2/U$, since the electronic
stabilization is less than $J$, but the singularity actually increases
\cite{sooshay2}; the GS energy in Eq.~(\ref{tre}) at $U = \Delta = 0$
 goes as $\delta ^2\ln {\delta}$ for alternating $t(1 \pm \delta)$
along the chain, while the GS of the spin chain with alternating
$J(1\pm \delta)$ goes  as $\delta ^{4/3}\ln {\delta}$
\cite{black}. Such considerations apply to Eq.~(\ref{uno})
in the covalent limit $\Gamma = U - 2\Delta >> t$ where,
as noted originally \cite{mcconnell,soosklein}, we have a
Heisenberg chain with $J = t^2/\Gamma$.

The charge gap is a recent topic and is expected to have a minimum at
NIT \cite{takada,qin,voit}.
We find finite minima in interacting systems with finite $N$.
As already noted, the polarizability
\cite{resta} and the charge stiffness in Eq.~(\ref{dodici}) and Fig. 8
give independent indications of a metallic GS at the NIT.
We consequently expect vanishing charge gap there.
The metal separating two insulating phases is extremely fragile:
not only is it restricted to a single point $\Delta _c(U)$,
but it is unconditionally unstable to dimerization.
Moreover, extending the model in Eq.~(\ref{uno}) to include
intersite e-e interactions or on-site phonons produces a
discontinuous NIT above some critical coupling, which excludes a
metallic phase even for a rigid lattice. By contrast, a metallic
 GS at the NIT of Eq.~(\ref{uno}) is fairly robust.
We have a simple half-filled band at $U = 0$ and a correlated metal
persists to arbitrarily large $U$. At the NIT,  $\Delta$
counterbalances $U$: the charge distributions DA and
D$^+$A$^-$ are almost degenerate and hence strongly mixed by
any finite $t$. A metallic state at NIT is not consistent
with finite triplet excitation there.

 The degeneracy of charge and spin excitations is characteristic 
of the NIT and appears already in the $t = 0$ limit of Fig. 1. 
Finite $\Delta$ completely spoils spin-charge separation at the NIT of
Eq.~(\ref{uno}). Vanishing $E_{ST}$ is closely linked to magnetic 
susceptibility of Hubbard or Heisenberg chains. 
Since a singlet can always be constructed from two 
triplets, vanishing $E_{SS}$ follows immediately and is associated 
with even-parity spin waves in $\Delta = 0$ systems with e-h symmetry. 
This symmetry is broken in Eq.~(\ref{uno}) or its extensions, and 
the CT excitation connecting the $A_1$ and $A_2$ GS is
 dipole allowed. Spin-charge separation is regained on the 
covalent side when the charge gap exceeds a few $t$, much as in 
Hubbard models for $U > t$: Exact separation requires infinite 
$U$, but $U > 4t$ suffices in practice.

To summarize, we have extended exact solutions of the modified Hubbard 
model in Eq.~(\ref{uno}) to larger systems, identified the NIT 
with symmetry crossovers in rings with periodic or antiperiodic 
boundary conditions, and found the charge density, excitation 
thresholds and susceptibilities at the NIT. 
We find a continuous NIT and 
tighten considerably the extrapolated limits for the infinite chain. 
Our results indicate a T=0K transition with vanishing singlet 
and triplet gaps on the covalent side, vanishing charge gap and metallic 
GS at the NIT, and finite singlet, triplet and charge gaps on the 
paired side, whose localized GS is confirmed.
We associate the NIT of 
the model in Eq.~(\ref{uno}) with a transition from a delocalized
(small $\Delta$) to  a localized (large $\Delta$) GS. 
Accurate analysis of the hamiltonian in Eq.~(\ref{uno}) 
is required to model valence transition in charge-transfer salts or 
metal oxides where long-range Coulomb interactions 
and e-ph coupling have to be considered explicitly.

\centerline { \bf {ACKNOWLEDGMENTS}}

One of us (A.P.) thanks A.Girlando for helpful discussions, and 
J.Voit for exchanging interesting correspondence on the subject
and for sharing information on unpublished work.
We gratefully acknowledge support for work at Princeton 
from the National Science Foundation through DMR-9530116 
and the MRSEC program under DMR-9400362, and, for work in Parma
from  the
 Italian  National Research Council (CNR)
within its ``Progetto Finalizzato Materiali Speciali per Tecnologie
Avanzate II'',   
and by the Ministry of University and of Scientific
and Technological Research (MURST).

\begin{table}
\caption{ Exact charge gap, $I - A$ in Eq.~(\ref{sei}), 
at the neutral-ionic transition  of the model in  
Eq. (1), for rings of $N$ sites, with $t = 1$, variable $U$ and 
$\Delta_c(N)$ at the crossover of 4n (4n+2) rings with periodic 
(antiperiodic) boundary conditions.}
\begin{tabular}{cccc}
$N$ & $U=2$ & 10 & $\infty$ (restricted basis)\\
\tableline
8 & 0.1893 & 0.7162 & 0.8486 \\
10 & 0.1803 & 0.6548 & 0.7614 \\
12 & 0.1718 & 0.6010 & 0.6909\\
14 & 0.1614 & 0.5555 & 0.6342\\
16 &        &        & 0.5874\\
18 &        &        & 0.5469\\
\end{tabular}
\end{table}

\begin{table}
\caption{Approximate ground states energy per dimer, Eq. (9), 
of the infinite chain and exact results for Eq. (1) with 
$N$ = 12, $t$ = 1, $\Gamma =\Delta  - \Delta _c(12,U)$ and 
$U$ = 10 and $\infty$ (the restricted basis).}
\begin{tabular}{cccc}
$\Gamma$ & $U = 10$, exact & $U = \infty$, exact & $U =\infty$ , Eq. (9)\\
\tableline
0.5 & -2.4706 & -2.3757 & -2.3148 \\
2.0 & -4.8129 & -4.7795 & -4.7871 \\
5.0 & -10.3848 &-10.3788 &-10.3814 \\
10.0 & -20.1981 & -20.1971 & -20.1975\\
\end{tabular}
\end{table}

\newpage
\vskip 15mm
\protect
\includegraphics [scale=0.45,angle=0]{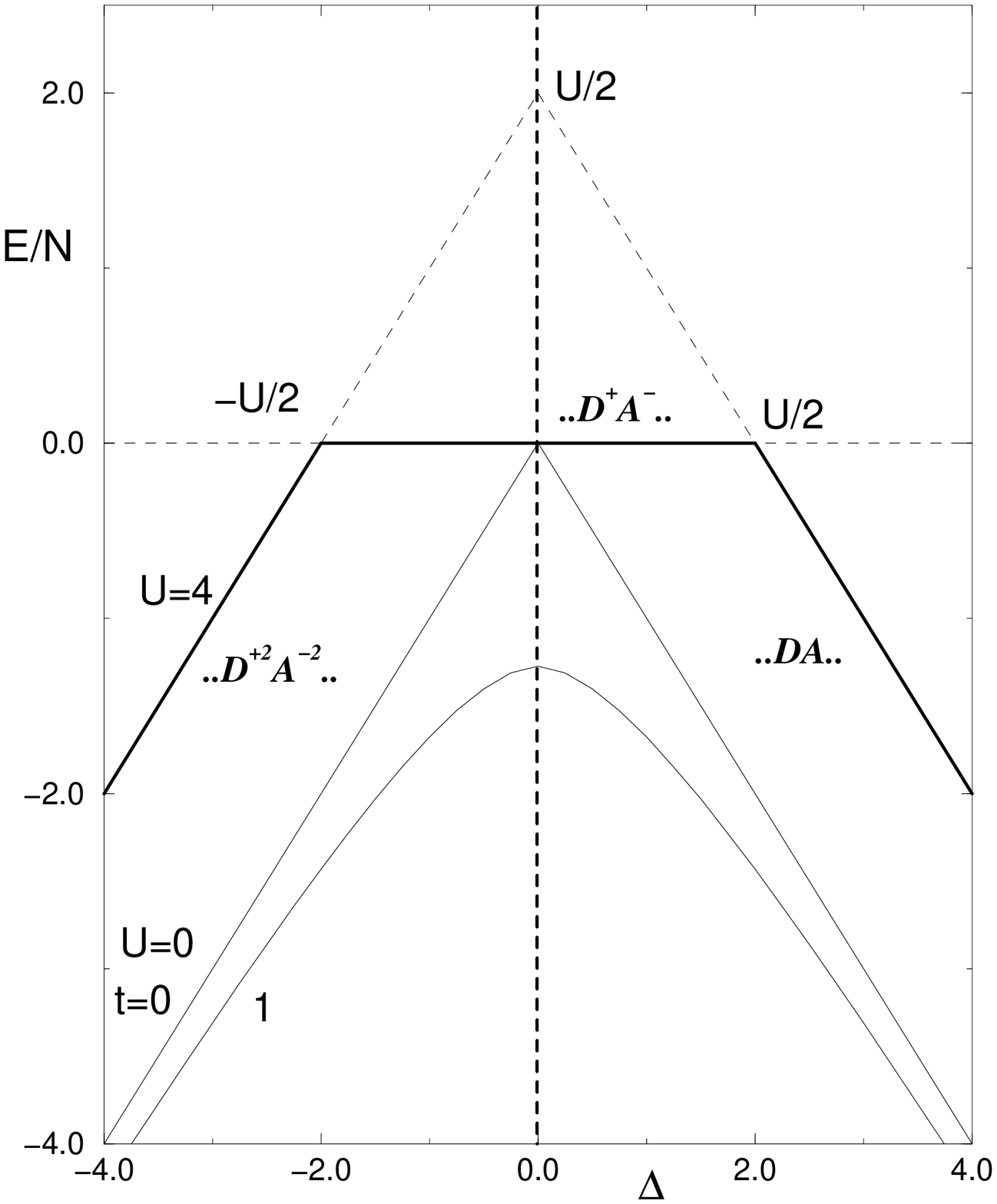}

\begin{figure}
\caption{ Ground-state energy per site, $E/N$, of the modified Hubbard model,
 Eq. (1), as a function of the site energy $\Delta$ for free ($U = 0$) 
or interacting ($U > 0$) electrons in the limit of no overlap ($t = 0$), 
with valence transitions at $\Delta = \pm  U/2$ in donor-acceptor stacks. 
The $t = 1$ curve for free electrons is Eq. (3).}
\end{figure}

\vskip 15mm
\protect
\includegraphics [scale=0.45,angle=0]{anna2.eps}

\begin{figure}
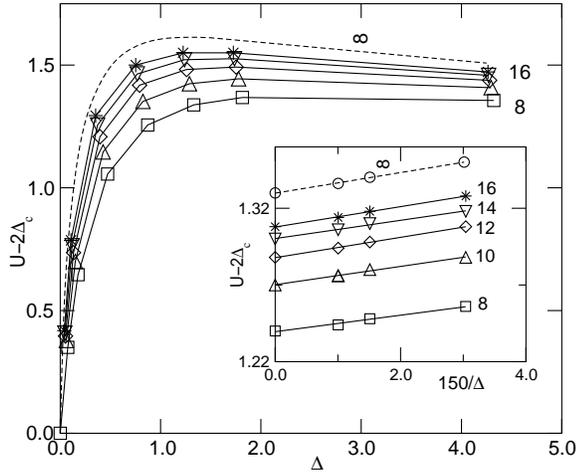

\caption{Ground state crossovers, $U(\Delta _c,N)$, of $N$-site modified 
Hubbard rings (Eq. (1)) with periodic and antiperiodic boundary conditions, 
respectively, for $N$ = 4n  and 4n+2. The dashed lines are 
$N \rightarrow \infty$  extrapolations discussed in the text. 
The inset shows the large-$\Delta$ behavior and the restricted basis 
at $\Delta \rightarrow \infty$.}
\end{figure}

\vskip 15mm
\protect
\includegraphics [scale=0.45,angle=-90]{anna3.eps}

\begin{figure}
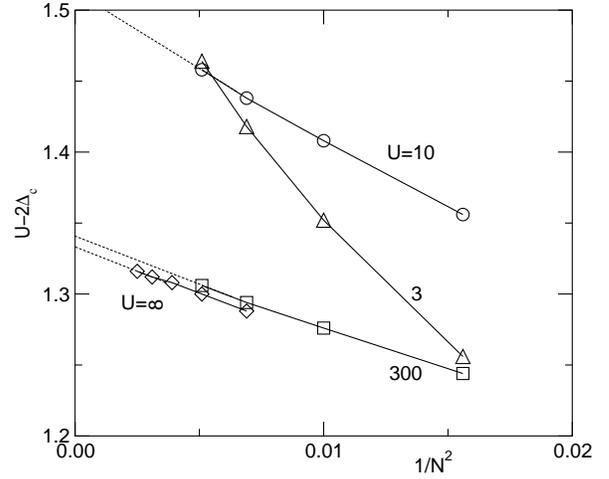

\caption{ Size dependence of the gs crossover between $N$ = 8 and 14 at 
$U/t$ = 3, 10, 300 for the full basis of Eq. (1) and up to $N$ = 20 
in the restricted basis with infinite $U$.}
\end{figure}

\vskip 15mm
\protect
\includegraphics [scale=0.35,angle=-90]{anna4.eps}

\begin{figure}
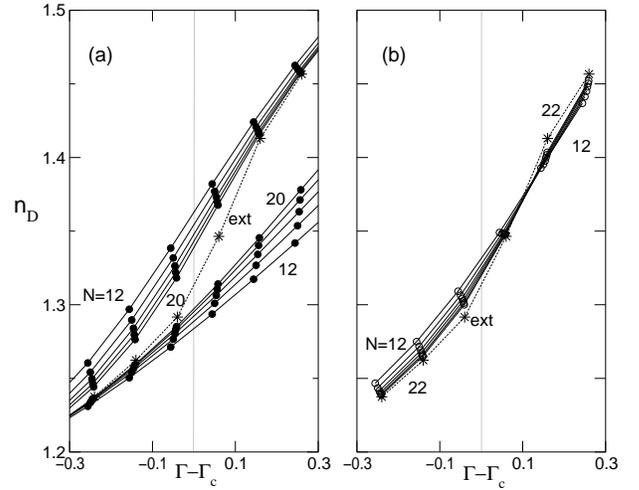

\caption{Ground-state electron density, $n_D$, of $N$-site modified 
Hubbard rings, Eq. (1), in the restricted basis. The boundary 
conditions in (a) produce symmetry crossover at $\Gamma _c(N)$, 
the vertical line, 
where $n_D$  increases discontinuously with $\Gamma$ 
and the smallest jump occurs for $N$ = 20. The boundary conditions in 
(b) with the same $\Gamma _c(N)$ result in continuous $n_D$ with 
increasing $\partial n_D/\partial \Gamma$  up to $N$ = 22. 
The stars are joint $N \rightarrow \infty$ extrapolations of 
(a) and (b) discussed in the text.}
\end{figure}

\vskip 15mm
\protect
\includegraphics [scale=0.40,angle=-90]{anna5.eps}

\begin{figure}
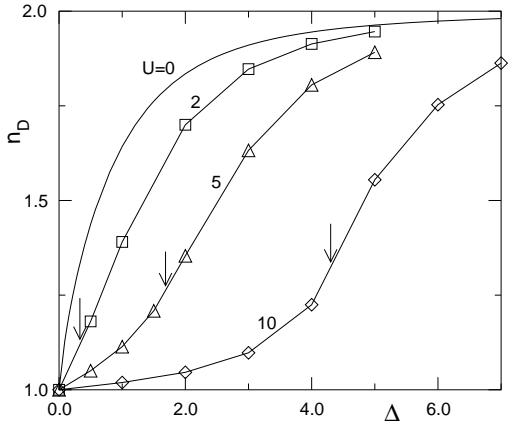

\caption{Ground-state electron density, $n_D$, of modified Hubbard 
models, Eq. (1), with $U/t$ = 0, 2, 5 and 10. The exact $U = 0$
 result is Eq. (4); $U > 0$ points are $N \rightarrow \infty$ 
extrapolation of $n_D$ based on the full basis up to $N = 16$; 
the arrows mark the neutral-ionic transition found as in Fig. 3.}
\end{figure}

\vskip 15mm
\protect
\includegraphics [scale=0.4,angle=-90]{anna6.eps}

\begin{figure}
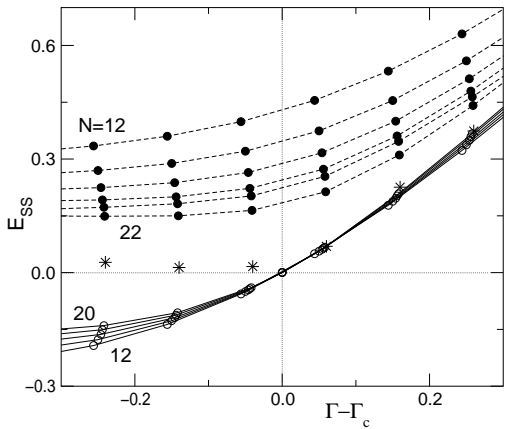

\caption{The singlet-singlet gap, $E_{SS}$, near the NIT of Eq. (1) 
up to $N = 22$ in the restricted basis. Boundary conditions leading 
to crossovers are shown as open circles and $|E_{SS}|$
 is the excitation for $\Gamma < \Gamma _c$.
 Boundary conditions without crossovers are shown as closed circles. 
The stars are joint   $N \rightarrow \infty$   extrapolations based on both.}
\end{figure}

\vskip 15mm
\protect
\includegraphics [scale=0.4,angle=-90]{anna7.eps}

\begin{figure}
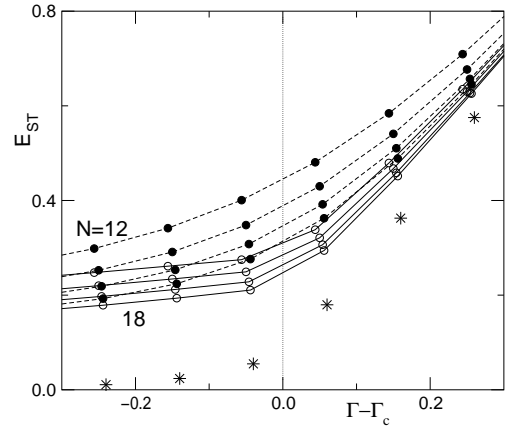

\caption{The singlet-triplet gap, $E_{ST}$, near the NIT of Eq. (1) 
up to $N = 18$ in the restricted basis. Open and closed circles refer 
to boundary conditions with and without crossovers, respectively, 
and stars are joint   $N \rightarrow \infty$     extrapolations based on both.}
\end{figure}

\vskip 15mm
\protect
\includegraphics [scale=0.4,angle=-90]{anna8.eps}

\begin{figure}
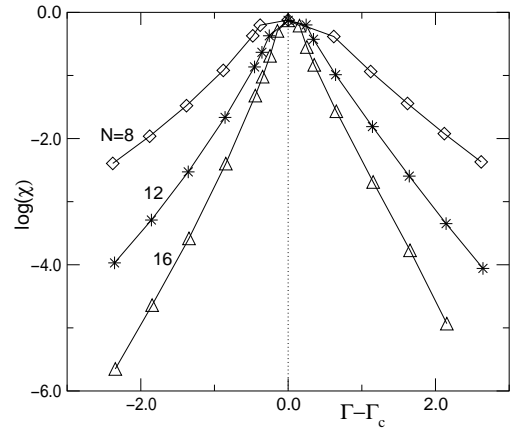

\caption{ Charge stiffness, $\chi_{cs} (\Gamma)$ in Eq. (12) and as discussed 
in the text, near the NIT of $N$-site model, Eq. (1), in the restricted basis;
  $\chi_{cs} (\Gamma _c)$ is $\sim 60$ \% of the free-fermion value.}

\end{figure}

\vfill
\eject


\begin{references}


\bibitem{mcconnell}  H.M. McConnell, B.M. Hoffman and R.M. Metzger, 
Proc.\ Natl.\ Acad.\ Sci.\ US \ {\bf 53}, 46 (1965);
P. L. Nordio, Z.G. Soos, and H.M. McConnell, 
Ann.\ Rev.\ Phys.\ Chem.\ {\bf 17}, 237 (1966).

\bibitem{soosklein} Z.G. Soos and D.J. Klein, in 
{\it Treatise on Solid State Chemistry}, Vol. III,
 N.B. Hannay, ed., Plenum, New York, (1976), p. 689.

\bibitem{strebelsoos} P.J. Strebel and Z.G. Soos, J. Chem.\ Phys. {\bf 53}, 
4077 (1970).

\bibitem{torrance} J.B. Torrance, Phys.\ Rev.\ Lett.\ {\bf 46}, 253 (1981); 
ibid. {\bf 47}, 1747 (1981)

\bibitem{girlando} A. Girlando,et al.  J. Chem.\ Phys.\ {\bf 79}, 1075 (1983).

\bibitem{egami} T. Egami, S. Ishihara and M. Tachiki, Science {\bf 261}, 
1307 (1993); T. Egami and M. Tachiki, Phys. \ Rev.\ B \ {\bf 49}, 8944 (1994).

\bibitem{soosmazu} Z.G. Soos and S. Mazumdar, Phys.\ Rev.\ B\ {\bf 18},
 1991 (1978).

\bibitem{gp86} A. Girlando and A. Painelli, Phys. \ Rev.\ B \ {\bf 34}, 
2131 (1986).

\bibitem{pg88} A. Painelli and A. Girlando, Phys. \ Rev. \  B \ {\bf 37}, 
5748 (1988); ibid. B \ {\bf 39}, 9663 (1989).

\bibitem{rice} M.J. Rice, Solid \  State  \ Commun. \ {\bf 31}, 93 (1979).

\bibitem{bibbie} A.Painelli and A. Girlando, J.\  Chem. \  Phys.\ {\bf 84},
 5655 (1986); R. Bozio and C. Pecile, in {\it Spectroscopy of 
Advanced Materials}, Adv. \ Spectrosc. Vol 19, R.J.H. Clark and R.E. Hester, 
eds., Wiley, New York, 1991, p.1.

\bibitem{pg87} A. Painelli and A. Girlando, J.\ Chem. \ Phys.\ {\bf 87}, 
1705 (1987).

\bibitem{sistematica} A. Girlando, A. Painelli and C. Pecile, 
Mol. \ Cryst.\ Liqu. \ Cryst.\ {\bf 120}, 17 (1985);
C. Pecile, A. Painelli, and  A. Girlando, ibid, {\bf 171}, 69 (1989).

\bibitem{pg92}  A. Painelli and A. Girlando, Phys. \ Rev. \  B \ {\bf 45}, 
8913 (1992).

\bibitem{nagaosa} N. Nagaosa and J. Takimoto, J.\ Phys.\ Soc.\ Japan \ 
{\bf 55},
 2735, 2747 (1986); N. Nagaosa, ibid. {\bf 55}, 2756 (1986).

\bibitem{avignon} M. Avignon, Phys.\ Rev.\ B\ {\bf 33}, 205 (1986); 
E.R. Gagliano, C.A. Balseiro and B. Alascio, Phys.\ Rev.\ B \ {\bf 37}, 
5697 (1988).

\bibitem{horovitz} B. Horovitz and J. Solyom,  Phys.\ Rev.\ B\ {\bf 35}, 
7081 (1987). 

\bibitem{nlo} A. Painelli, Chem.\ Phys.\ Lett.\ {\bf 285}, 352 (1998).

\bibitem{torino} A. Painelli and A. Girlando, in {\it Interacting Electrons 
in Reduced Dimension}, D. Baeriswyl and D.K. Campbell, eds., NATO ASI B 213, 
Plenum, New York (1989)  p.189; D. Baeriswyl, D.K. Campbell and S. Mazumdar, 
in {\it Conducting Polymers}, H. Kiess, ed., Springer-Verlag, Heidelberg 
(1992), p. 7.

\bibitem{handbook} T.E. Skotheim, R.L. Elsenbaumer and J.R. Reynolds, eds., 
{\it Handbook of Conducting Polymers}, 2nd ed., Marcel Dekker, New York (1998).

\bibitem{sooshay} Z.G. Soos and G.W. Hayden, in 
{\it Electroresponsive Molecular and Polymeric Systems}, T.E. Skotheim, ed., 
Marcel Dekker, New York (1988), p. 197.

\bibitem{nasu}  K. Nasu, ed., {\it Relaxations of Excited States 
and Photo-Induced Structural Phase Transitions}, Springer series in 
Solid-State Sciences 124, Springer-Verlag, Heidelberg (1997).

\bibitem{sbm} Z.G. Soos, S.R. Bondeson, and S. Mazumdar, 
Chem.\  Phys.\ Lett.\ {\bf 65}, 331 (1979).

\bibitem{resta} R. Resta and S. Sorella, Phys.\ Rev.\ Lett.\ {\bf 74}, 
4738 (1995); ibid. {\bf 82}, 370 (1999).

\bibitem{fabrizio} M. Fabrizio, M.O. Gogolin and A.A. Nersesyan, 
Phys. \ Rev.\ Lett.\ {\bf  83}, 2014 (1999).

\bibitem{ortiz} G. Ortiz, P. Ordejon, R.M. Martin, and G. Chiappe, 
Phys.\ Rev.\ B\ {\bf 54}, 13515 (1996).

\bibitem{tosatti} N. Gidopoulos, S. Sorella and E. Tosatti, 
Eur.\ Phys.\ J.\ B\ {\bf 14}, 217 (2000).

\bibitem{caprara} S. Caprara, M. Avignon, O. Navarro,  Phys. \ Rev. \  B \ 
{\bf 61}, 15667  (2000).

\bibitem{takada} Y. Takada, M. Kido, cond-mat/0001239.

\bibitem{qin} S. Qin, et al. cond-mat/0004162.

\bibitem{vb} Z.G. Soos and S. Ramasesha, in {\it 
Valence Bond Theory and Chemical Structure}, D.J. Klein and N. Trinajstic, 
eds., Elsevier, New York  (1990), p. 81; 
G. Wen and Z.G. Soos, J.\ Chem.\ Phys.\ {\bf 108}, 2486 (1998).

\bibitem{lieb}  E.H. Lieb and F.Y. Wu, Phys.\ Rev.\ Lett.\ {\bf 25}, 
1445 (1968).

\bibitem{ovchi} A.A. Ovchinnikov, Sov.\ Phys.\ JETP\ {\bf 30}, 1160 (1970).

\bibitem{taka} M. Takahashi, Prog.\ Theor.\ Phys.\ {\bf 42}, 1098 (1969); 
ibid. {\bf 43}, 1619 (1970).

\bibitem{mclac} A.D. McLachlan, Mol.\ Phys.\ {\bf 2}, 276 (1959); 
O.J. Heilmann and E.H. Lieb, Trans.\ N.Y.\ Acad.\ Sci.\ {\bf 33}, 116 (1971); 
S.R. Bondeson and Z.G. Soos, J.\ Chem.\ Phys.\ {\bf 71}, 380 (1979).

\bibitem{skh} Z.G. Soos, S. Kuwajima and R.H. Harding, J.\ Chem.\ Phys.\
{\bf  87}, 1705 (1986).

\bibitem{kohn} W. Kohn, Phys.\ Rev.\  A\ {\bf 133}, 171 (1964).

\bibitem{ss} B.S. Shastry and B. Sutherland, Phys.\ Rev.\ Lett.\ {\bf  65}, 
243 (1990); C.A. Stafford, A.J. Millis and B.S. Shastry, Phys.\ Rev.\ B\
{\bf 43}, 13660 (1991); R.M. Fye, M.J. Martin, D.J. Scalapino, 
J. Wagner and W. Hanke, Phys.\ Rev.\ B\ {\bf 44}, 6909 (1991).

\bibitem{sap} Z.G. Soos, Y. Anusooya-Pati and S.K. Pati, J.\ Chem. \ Phys.\
{\bf  112}, 3133 (2000).

\bibitem{tsiper} E.V. Tsiper and A.L. Efros, J.\ Phys.:\ Condens.\ Matter\
{\bf  9}, L561 (1997).

\bibitem{majumdar} C.K. Majumdar, and D.K. Ghosh, J.\ Math.\ Phys.\
{\bf 10}, 1388 (1969).

\bibitem{bray} J.W. Bray, L.V. Interrante, I.S. Jacobs and J.C. Bonner, 
in {\it Extended Linear Chain Compounds}, Vol. 3, J.S. Miller, ed. Plenum, 
New York (1983), p. 353.

\bibitem{sooshay2} Z.G. Soos and G.W. Hayden, Mol.\ Cryst.\ Liqu.\ Cryst.\
{\bf  160}, 421 (1988).

\bibitem{black} J.L. Black and V.J. Emery, Phys.\ Rev.\ B\ {\bf 23}, 
429 (1981).

\bibitem{voit} J.Voit, M. Nakamura, private communication. 


\end{references}
\end{document}